\documentclass[pre,twocolumn,floatfix,showkeys,showpacs]{revtex4}

\usepackage{amssymb,amsfonts,amsmath}
\usepackage{graphicx}

\newcommand \beq {\begin{equation}}
\newcommand \bea {\begin{eqnarray} \nonumber }
\newcommand \eeq {\end{equation}}
\newcommand \eea {\end{eqnarray}}
\newcommand{\beqa}{\begin{eqnarray}}
\newcommand{\eeqa}{\end{eqnarray}}

\newcommand{\avg}[1]{\langle{#1}\rangle}

\newcommand{\rme}{\text{e}}

\newcommand{\aref}[1]{(\ref{#1})}
\newcommand{\logit}[0]{{\rm logit \,}}

\begin{document}

\title{Fitness in time-dependent environments includes a geometric
 contribution}

\author{Sorin T\u{a}nase-Nicola}
\affiliation{Department of Physics\\ Emory University, Atlanta, GA
  30322}
\email{sorintan@physics.emory.edu}
\author{Ilya Nemenman}
\affiliation{Departments of Physics and Biology and\\
    Computational and Life Sciences Initiative\\ Emory University,
    Atlanta, GA 30322}
\email{ilya.nemenman@emory.edu}

\begin{abstract}
  Phenotypic evolution implies sequential fixations of new genomic
  sequences. The speed at which these mutations fixate depends, in
  part, on the relative fitness (selection coefficient) of the mutant
  vs.\ the ancestor.  Using a simple population dynamics model we show
  that the relative fitness in dynamical environments is not equal to
  the fitness averaged over individual environments. Instead it
  includes a term that explicitly depends on the sequence of the
  environments. This term is geometric in nature and depends only on
  the oriented area enclosed by the trajectory taken by the system in
  the environment state space. It is related to the well-studied
  geometric phases in classical and quantum physical systems. We
  discuss possible biological implications of these observations,
  focusing on evolution of novel metabolic or stress-resistant functions.
\end{abstract}
\maketitle

\section{Introduction}

Organisms react to long-term changes in environmental conditions by
sequential fixation of new genome sequences, mostly corresponding to
increasingly more adapted phenotypes. However, often environmental
changes are faster than the characteristic time for mutation-selection
cycles needed to evolve an optimal phenotype. In such cases, depending
on the structure and time scales of the fluctuations, a dynamic
environment creates dynamic fitness landscapes \cite{Mustonen09},
promotes sensing \cite{Kussell05}, modularity \cite{Kashtan05,Sun07},
switching \cite{Beaumont09}, and can change the speed of adaptation
\cite{Kashtan07, Mustonen08}.

The effect of fluctuating selection and/or population size on the
population-genetics dynamics have been extensively studied over the
years \cite{Gillespie94,Mustonen08}, starting with the introduction
of the concept of adaptive topography by Wright \cite{Wright1932a}.
More recently, the evolutionary dynamics of density regulated
populations in fluctuating environments has been elucidated in 
ecologically realistic models \cite{MacArthur01,Heckel80,Lande09}.
These bridge the gap between the classical population dynamics
exhibiting very diverse responses to fluctuating environments
\cite{Cushing86,Namba84} and classical population genetics models.
However, a complete understanding of the effect of fluctuations on
population and evolutionary dynamics has not been achieved yet.

Some of the relevant parameters describing evolutionary response of a
population to a changing environment are the rate at which new
genotypes are created (mutation rate), the relative fitness of new
phenotypes, and the total population size. We concentrate on the case
of environments changing on scales longer than an individual's
lifetimes. This is relevant, in particular, for bacterial populations
confronted with daily environmental changes (natural or artificial)
\cite{Cooper10}, for longer-living organism affected by seasonal
variations, or for pathogens experiencing transmission, uncontrolled
growth in a new host, and then effects of the host immune system. For
example, in the now-classic long-term {\em E.\ coli} evolution
experiment \cite{Lenski11}, bacterial cultures are diluted daily, and
the environment (i.e., cell growth and death rates) changes during
dilution events and between them due to depletion of resources, cell
density growth, and cell-to-cell interactions. These experiments are a
great model to study clonal competition
\cite{Barrick2009}. Interestingly, the number of accumulated
beneficial mutations is relatively small, considering that every
single point and many possible double mutations have happened
thousands of times in the 25-year history of the experiment.  This
discrepancy is likely largely accounted for by strong bottlenecks at
dilution times, when most new mutations disappear by chance.  However,
all clones, even beneficial ones, experience additional huge
fluctuations in their reproductive rates during the course of the
experiment. It remains to be seen if such fluctuations can contribute
to the slowing down of the evolutionary adaptation as well.

In this article, we make a step in this direction by studying effects
of fluctuating environments (represented by birth and death rates) on
the effective selection coefficient. Using analytic and computational
tools, we investigate a model of a heterogeneous population (a
background strain and a newly emergent mutant) under the assumption
that the time scales of the clonal frequency dynamics on the one hand
and the environment fluctuation on the other are both much larger than
the division time, but not necessary well separated from each other.
We start by showing that the selection coefficient in an infinitely
slowly changing environments is given by a time-average of static
selection coefficients corresponding to each environment. However, for
environments varying at a slow but finite rate, such time-average is
not the whole story. A new contribution emerges. For example, in a
cyclically oscillating environment, this contribution to the selection
coefficient is {\em independent of the
  speed} of variation and depends only on the {\em sequence of
  environments} visited during each cycle. The contribution is
non-zero only for nontrivial coupling between the environment and the
population dynamics, represented as a multi-dimensional trajectory in
the space of birth and death rates. The contribution changes sign when
the sequence of the visited environments is reversed. It is largely
independent on the speed of the dynamics. Finally, it scales
quadratically with the amplitude of the environmental fluctuations. In
other words, the contribution is {\em geometric} in nature. We believe
that this has not been noticed before in the context of population
dynamics.

We will focus on the deterministic approximation to population
dynamics. Geometric effects are well-known for slowly changing
deterministic dynamical systems \cite{Shapere1988,Sinitsyn09}. While
evolutionary dynamics of a population driven together by forces of
mutation, drift and selection cannot be accurately described
deterministically, we believe that our model is meaningful even for a
stochastic case for large population, low mutation rate, and strong
selection. Indeed, the recent observation that stochastic dynamical
systems are also subject to geometric corrections suggests that
deterministic vs.\ stochastic treatment of population dynamics is not
crucial for the phenomenon \cite{Sinitsyn07a,Sinitsyn07b}.

In what follows, we develop our results in a relatively simple two
species population model with bilinear, symmetric competition, which
we believe is general enough to capture the main effects of
fluctuations for a large class of related models. We first solve the
system in the limit of small differences between the birth and the
death rates of the competing species. We derive expressions for the
selection coefficient in the limit of stationary, very slowly
continuously, and infrequently discontinuously varying environments.
The selection coefficient for arbitrary time scale of the environment
fluctuations can be derived then using a perturbative approach.

\section{Model}

Let $x_{i}$ be the number of individuals of genotypes $i,\, i=1,2,$ in
a large asexual population. We assume that $x_i\gg 1$, so that
demographic (phenotypic) fluctuations and random genetic drift can be
neglected. We refer to $x_1$ as an ancestral phenotype, and to $x_2$
as a mutant. The competition between the two is described by a driven
two-dimensional Lotka-Volterra (logistic) model \cite{Namba84,Cushing86}
\bea
\dot x_1& =& x_1\left[b_1(t)-d_1(t)(x_1+x_2)\right],\\
\dot x_2 &=& x_2\left[b_2(t)-d_2(t)(x_1+x_2)\right].
\label{eq:mot}
\eea Here $b_i(t)$ represents the birth rates, and $d_i(t)$
parameterize the death rates for each of the genotypes. Generally, all
parameters are time dependent.

Following classical models of ecological population genetics, we view
our model as a particular form of the more general dynamics. Defining
the total population size, $x(t)=x_1(t)+x_2(t)$, we write \bea
\dot x_1 & =& x_1g_1(x,t),\\
\dot x_2 &=& x_2g_2(x,t).
\label{eq:megm}
\eea Here $g$ is the generalized growth rate. For this system of
equations to represent the dynamics of a realistic self-sustaining
population, $g_i(x)$ must be negative for large $x$, and it must have
at least one zero. Our approach applies to a very general subset of
such growth rate functions provided that the system,
Eq.~(\ref{eq:megm}), has exactly one fixed point on each of the axes
$x_i=0$ in addition to the trivial unstable extinction point $(0,0)$.

One traditionally takes \cite{Lande09}
\begin{equation}
g_i= r_1(t)\left[1-\frac{f(x)}{f(K_i(t))}\right],
\label{eq:eco}
\end{equation}
where $r_i$'s are the intrinsic maximum growth rates of each genotype,
if unconstrained by limited resources. The terms $r_i
{f(x)}/{f(K_i(t))}$ represent the reduction of these rates due to
competition for resources. This reduction depends only on the total
population size $x(t)$ and on $K_i$, which are stable total
populations of the isolated phenotypes $i$ supported by stationary
resource-limited environments. $K$'s are referred to as the carrying
capacities. Our approach applies for any non-negative, monotonously
increasing $f(x)$, as explained above. However, for simplicity, we now
concentrate on $f(x)=x$. In this case, the competition is linear and
symmetric, and the simple Lotka-Volterra model \aref{eq:mot} is
recovered with $b_i(t)=r_i(t)$ and $d_i(t)={r_i(t)}/{K_i(t)}$.

We are interested in modeling competition of the ancestral genotype
with the mutant one. The two are very close in the genotype space,
essentially one mutation away.  Since mutation effects are, in
general, small \cite{Barrick2009}, we assume that the differences
between $g_1$ and $g_2$ are also small, \beq \left|
  \frac{g_1(x,t)-g_2(x,t)}{g_1(x,t)+g_2(x,t)}\right| \leq \epsilon \ll
1.
\label{eq:smc}
\eeq This corresponds to small differences in the parameters $b_i$,
$d_i$, $r_i$, $K_i$. We assume this from now on. In particular, it is
possible that differences between the mutant and the ancestor
parameters at any particular time are much smaller than the variations
of each of the parameters over time.

\section{Preliminaries}

In order to determine the conditions under which the mutant, initially
present in small numbers relative to the ancestor, invades the
population, we explicitly integrate the model, Eq.~(\ref{eq:mot}).  We
write the dynamics of the total population size $x=x_1+x_2$: \beq \dot
x= x \left[\left(\frac{b_1(t) x_1}{x}+\frac{b_2(t)
      x_2}{x}\right)-\left(\frac{d_1(t) x_1}{x}+\frac{d_2(t)
      x_2}{x}\right)x\right]. \eeq To the zeroth order in
$\epsilon\ll1$, this does not depend on the individual values $x_1$
and $x_2$: \beq \dot x= x \left[b(t)-d(t)x\right]+O(\epsilon), \eeq
where we have defined \beq b(t)=\frac{b_1(t)+b_2(t)}{2},\quad
d(t)=\frac{d_1(t)+d_2(t)}{2}.  \eeq We also define \beq p=\frac{
  x_2}{x_1+x_2}, \eeq the fraction of the mutant in the whole
population. This obeys \bea \dot
p=p(1-p)\left\{\left[b_2(t)-b_1(t)\right]-\left[d_2(t)-d_1(t)\right]x\right\}.
\eea The model then reduces to \bea
\dot x&=&x\left[b(t)-d(t)x\right], \\
\dot p&=&p(1-p)\left[\delta b(t) -\delta d(t) x\right],
\label{eq:model2}
\eea where we have used the notation $\delta (b,d)$ for small (order
$\epsilon$) time dependent differences between the corresponding
mutant and ancestral rates. To simplify the notation, for any pair of
parameters $({\cal{P}}_1,{\cal{P}}_2)$ describing the ancestor and the
mutant, we write ${\cal{P}}=\left({\cal{P}}_1+{\cal{P}}_2\right)/2$,
and $\delta {\cal{P}}={\cal{P}}_2-{\cal{P}}_1$. In addition we always
assume $\left|{\delta \cal{P}}/{ \cal{P}}\right| =O(\epsilon) \ll1$.

To the zeroth order in $\epsilon$, the dynamics of the total
population size defined by
Eqs.~\aref{eq:mot} is now uncoupled from the dynamics of the mutant fraction\beq
x(t)=\frac{x(0)\rme^{\int_0^t \!\! d\tau b(\tau) }}{1+x(0) \int_0^t  \!\! dt'  d(t')\rme^{\int_0^{t'}\! \!d \tau b(\tau) }}.
\label{eq:xsol}
\eeq
Due to the small variation assumption, Eq.~\aref{eq:smc}, $p(t)$
changes on time scales much longer than $x(t)$. On these time scales,
$x(t)$ converges to a unique (up to the first order in $\epsilon$)
attractor $x_{\rm{a}}(t)$, independent of the initial conditions,
\beq
x_{\rm{a}}(t)=\frac{1}{ \int_{-\infty}^t \!\!dt'\,  d(t')\,\rme^{\int_t^{t'}\!\!d\tau b(\tau) }}.
\label{eq:x}
\eeq
Then the slower dynamics of   $p$ is 
\beq
\logit p(t) = \logit p(0)+
\int_0^t \!\!d\tau\,\left[\delta b(\tau) -\delta d(\tau) x(\tau)\right],
\label{eq:solp}
\eeq where $\logit p=\log p -\log(1-p)$. The obvious first lesson from
this equation is that the clone with the largest average growth
rate, $\langle g_i\rangle\ge \frac{1}{\cal{T}}\int_0^{\cal{T}} \! dt
\,\left[b_i(t)-d_i(t)x(t)\right] $ for some large $\cal{T}$, will have
an advantage.

\section{Selection coefficient}
For coefficients varying periodically with a period $T$, we write
for the logarithmic change of the mutant-to-ancestor ratio, $\logit
p$, over time ${\cal T}\gg T$,
\bea
\Delta({\cal T})&\equiv& \logit p({\cal T})-\logit p(0)\\
&=&{\cal T} \frac{\int_0^T\!d \tau\,\left[\delta b(\tau) -\delta d(\tau)
      x(\tau)\right]}{T}\equiv s {\cal T},
\label{eq:solps}
\eea where the last equality defines the selection coefficient, $s$.
It is the sign of $s$ that decides the stability of the fixed points
$p=1$ and $p=0$. For example, for $s>0$, $p=0$ is unstable, and the
mutant phenotype invades the population towards a stable fixed point
$p=1$.

In a constant environment, and for $\epsilon \ll 1$,
the selection coefficient $s$ can be rewritten in terms of the
ecological parameters defined in Eq.~(\ref{eq:eco})
\beq
s \approx r   \frac{\delta K}{K}.
\label{eq:scon}
\eeq We have a classical result that selection favors phenotypes with
larger carrying capacities (larger $K_i$) independent of the magnitude
of the intrinsic growth rates $r_i$ \cite{MacArthur01,Heckel80}. To
derive this, we rely on the fact that the total population is given at
all times by $K$, and it is independent of the frequency of the mutants in
the population.

In this paper, we are interested in the values of the selection
coefficient for temporally varying environments. As a consequence, the
selection coefficient is now given by the interaction between several
varying quantities. To simplify the discussion, we focus on limiting
cases of large time scale separation between the environment
fluctuations and individual lifetimes.

In the regime of infinitely fast environmental fluctuations, for
${\cal T}\to\infty$, we approximate the general
driven model, Eq.~\aref{eq:megm}, as
\bea
\dot x_1& =& x_1 \avg{g_1(x)}_T,\\
\dot x_2 &=& x_2 \avg{g_2(x)}_T.
\label{eq:motga}
\eea We assume here that the environment variation attains a well
defined, constant average for every state $(x_1,x_2)$. We denote this
by $\avg{\dots}_T$, where the subscript $T$ stands for averaging over
a period. We assume that $x$ does not change appreciably over this
time. For the specific case of the Lotka-Volterra model, the selection
coefficient for fast fluctuations, $s_f$, can be computed using the
formula for the constant case, Eq.~\aref{eq:scon}, keeping in mind
that one has to use the average values of the relevant
coefficients: \beq s_f=\delta \avg{r} -\delta\avg{r/K}\, \frac{\avg{
    r} }{ \avg{r/K} }. \eeq

In the opposite limit of an infinitely slow parameter variation, the
total population is equal to the carrying capacity at all times,
$x(t)=K(t)$. In this case, the quasi-stationary (qst) selection
coefficient $s$ is \beq s_{\rm qst}=\frac{1}{T} \int_0^T \!dt\,r(t)
\frac{\delta K(t)}{K(t)} =\frac{1}{T} \int_0^T \!dt\,s(t), \eeq where
the period $T$ is much longer than the individual's lifetime. This
allows for a proper average to be attained.

In both limits, the sign of the selection coefficient does not depend
on the average carrying capacity \cite{Heckel80,Lande09}. Indeed, it
is possible to have a slowly varying environment, in which
the mutant has, on average, a larger carrying capacity but a lower
fitness. In both limits, the selection coefficient becomes independent
of the speed of environmental variations, and it is symmetric with
respect to time reversal for the driving parameters.

\section{Continuous, deterministic, oscillatory environments}

We now proceed to a more realistic case of an environment fluctuating
slowly, but not infinitely slowly, compared to an individual's
lifetime. This condition allows us to derive a perturbative
approximation for the selection coefficient valid when $b(t),r(t) \gg
\frac{1}{T}$ are satisfied at every $t$. Our approximation is based on
a simplified solution for the dynamics of the total population size
$x_{\rm a}(t)$, Eq.~\aref{eq:x}. By making a variable change
$y(t)=\int_0^t\!d\tau\, b(\tau) $, we write \beq x_{\rm
  a}(y)=\frac{1}{ \int_{-\infty}^y\! dz \frac{d(z)}{b(z)}
  \rme^{-(y-z)}}=\frac{1}{ \int_{-\infty}^y \!dz\frac{1}{K(z)}
  \rme^{-(y-z)}}. \eeq In the limit of slow environmental changes,
the carrying capacity $K(y)$ varies slowly, and the integral in the
denominator is dominated by the value of $1/K(z)$ around $z=y$. In
this regime, \beq \frac{1}{K(z)} \simeq
\frac{1}{K(y)}-\frac{K'(y)}{K^2(y)}(z-y) \quad \text{for} \quad
(y-z)\ll y.
 \label{eq:aK}
\eeq

Using Eq.~\aref{eq:aK}, we now derive an approximation for the total
population trajectory $x_{\rm a}$ valid in the qst regime. We denote
it as $x_{\rm qa}$,
\beq
x_{\rm qa}(t)\simeq K(t)-\frac{K'(t)}{r(t)}.
\label{eq:ax}
\eeq This solution represents the correction to the quasi-stationary
result $x_{\rm qst}(t)=K(t)$ as a first order perturbation in the
small ratio between the rate of change of the environment and the
typical rate of change of the total population. Note that the
approximation is consistent with the intuition that the instantaneous
total population falls behind the instantaneous carrying capacity.

The selection coefficient can be expressed now as \beq s=s_{\rm
  qst}+s_{\rm geom}, \eeq where \beq s_{\rm
  geom}=\frac{1}{T}\int_0^T\!d\tau\,\left[ \frac{\delta
    r(\tau)}{r(\tau)} -\frac{\delta K(\tau)}{K(\tau)}
\right]\frac{K'(\tau)}{K(\tau)} \label{eq:s} \eeq is a {\em geometric}
contribution to the selection rate. The geometric nature of this term
can be better understood if we express the change in the
mutant-to-ancestor ratio as \beq \Delta ({\cal T})=s_{\rm qst} {\cal
  T} + \Delta_{\rm geom} ({\cal T}). \label{eq:J} \eeq We note that,
for any reparameterization of time, $\lambda=\lambda(t)$, $\Delta_{\rm
  geom}$ can be written in a very similar form \beq \Delta_{\rm
  geom}({\cal{T}})=\int_0^{\Lambda({\cal{T}})}\!\!d\lambda\, \left[
  \frac{\delta r(\lambda)}{r(\lambda)} -\frac{\delta
    K(\lambda)}{K(\lambda)} \right]\frac{K'(\lambda)}{K(\lambda)},
\label{eq:I}
\eeq which emphasizes that it depends on the trajectory itself,
$r_{1,2}(\lambda)$, $K_{1,2}(\lambda)$, rather than on how this
trajectory is traversed. As any closed contour integral expression,
this expression can be transformed into a surface integral over any 2D
domain bounded by the trajectory $[r_1(t),r_2(t),K_1(t),K_2(t)]$ in
the parameter space. In particular, using variables \beq {\cal
  X}=\delta \log \frac{r}{K},\quad {\cal Y}=\log{K} \eeq and the
Stokes theorem, we can equate $\Delta_{\rm geom}({\cal T})$ with the
oriented area bounded by the trajectory for times $t\in (0,{\cal T})$
in the plane $(\cal{X},\cal{Y})$.

\begin{figure}[t]
\centerline{\includegraphics[width=8cm]{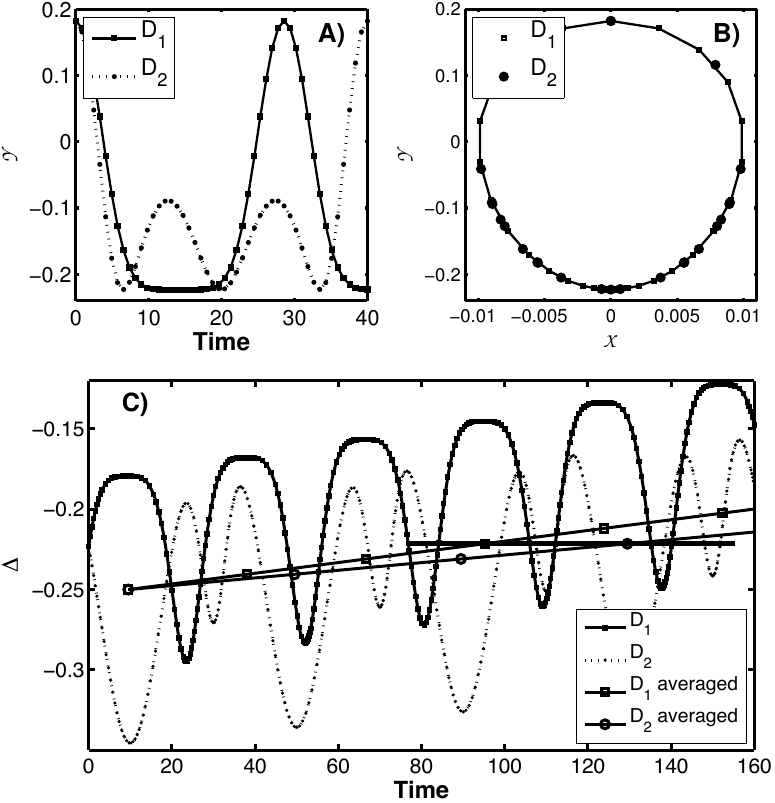}}
\caption{Mutant fraction as a function of time for two sample
  environment trajectories $({\cal X}(t), {\cal Y}(t))$:
  ${\cal{X}}_1(t)=0.02 \sin\left[\omega_1 t+\sin(\omega_1
    t)\right],\quad {\cal{Y}}_1(t)=1+0.1 \cos\left[\omega_1
    t+\sin(\omega_1 t)\right]$ and ${\cal{X}}_2(t)=0.02
  \sin\left[\omega_2 t+2.5 \sin(\omega_2t)\right],\quad
  {\cal{Y}}_2(t)=1+0.1 \sin\left[\omega_2 t+2.5 \sin(\omega_2
    t)\right]$ where $\omega_1/1.4=\omega_2=2 \pi/40$.  (A) The two
  trajectories for ${\cal Y}=\log K$ are shown; the first has the
  frequency 1.4 times the second, and the second reverses twice
  before completing the full cycle. (B) Nonetheless, the shapes of the
  trajectories $({\cal X}(t), {\cal Y}(t))$ are the same for both
  examples. (C) Instantaneous and one-period-averaged mutant fractions
  for both trajectories. The average growth of $\Delta$, given
  completely by a geometric term, is linear. The slopes of the two
  curves are different by exactly 1.4, so that $\Delta$ is only
  dependent on the number of elapsed periods. This is indicated by the
  horizontal line connecting the two averages delayed by the same
  number of periods. Thus the geometric contribution to the mutant
  fraction depends only on the shape of the contour in the parameter
  space and on the number of cycles, but is independent of the speed
  of the trajectory traversal.  \label{fig1} }
\end{figure}

\begin{figure}[t]
\centerline{\includegraphics[width=8cm]{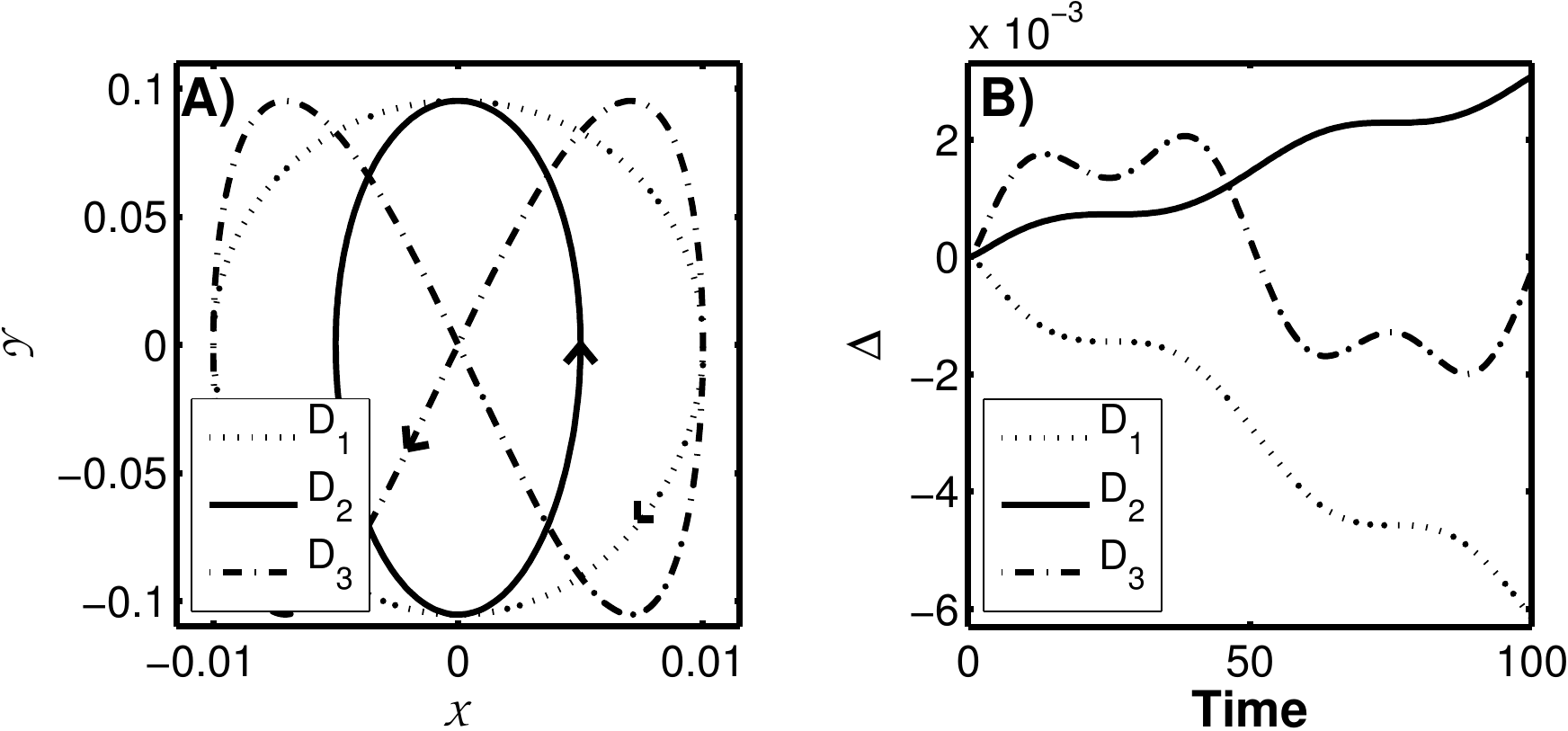}}
\caption{Illustration of the geometric nature of the mutant fraction
  dynamics. (A) Three different trajectories $({\cal X}(t), {\cal
    Y}(t))$: ${\cal{X}}_1(t)=-0.02 \cos(\omega t),\quad
  {\cal{Y}}_1(t)=1+0.1 \sin(\omega t)$, ${\cal{X}}_2(t)=0.01 \cos(\omega
  t),\quad {\cal{Y}}_2(t)=1+.1 \sin(\omega t)$, ${\cal{X}}_3(t)=0.02
  \cos(\omega t),\quad {\cal{Y}}_3(t)=1+0.1 \sin(2\omega t)$ where
  $\omega=2 \pi/100$. The second trajectory (solid line) encloses
  exactly half the area of the first (dotted line), and the two are
  traversed in opposite directions. The oriented area enclosed by the
  third trajectory (dash-dotted) is zero. (B) The average mutant
  fraction change for the first trajectory is equal to the oriented
  area and is, therefore, twice that for the second one, and in the
  opposing direction. The quantity is zero for the third trajectory. \label{fig2}}
\end{figure}

In other words, $\Delta_{\rm geom}$ is a truly geometric term in the
spirit of geometric phases in quantum or classical mechanics
\cite{Shapere1988,Sinitsyn09}. The geometric nature of the change in
the population composition over long times, Eq.~\aref{eq:I}, is {\em
  the main result} of the paper. It allows us to make important
macroscopic predictions about the population dynamics that will hold
generally irrespective of the microscopic details of the model. First,
the geometric changes in the relative fraction of the mutant depend on
the {\em sequence} of the environmental states in addition to their
identity: same environmental states may have very different effects
depending on the order in which the states are visited. At an extreme,
a reversal of the order (time-reversal) would change the sign of the
geometric contribution, which may make a deleterious mutation
advantageous, and vice versa. To our knowledge, such dependence of the
effective selection coefficient on the sequence of the environmental
states has not been noticed before in population biology. Second, the
contribution to $\Delta_{\rm geom}$ depends only on the {\em oriented
  area} covered in the parameter space (and thus, in particular, on
the number of periodic oscillations), but not on the speed of
traversal of the trajectory. Figure~\ref{fig1} illustrates these
features: even when the environmental dynamics involves backtracking,
the overall contribution per period still does not change. The
dependence on the area in the parameter space also suggests that the
geometric contribution scales as the square of the fluctuation
amplitudes. Finally, to achieve a nonzero area, more than one
parameter must be changing, and they must change incoherently. We
illustrate some of these features in Fig.~\ref{fig2}

\section{Switching among discrete environment states} 

\begin{figure}[th]
\centerline{\includegraphics[width=6cm]{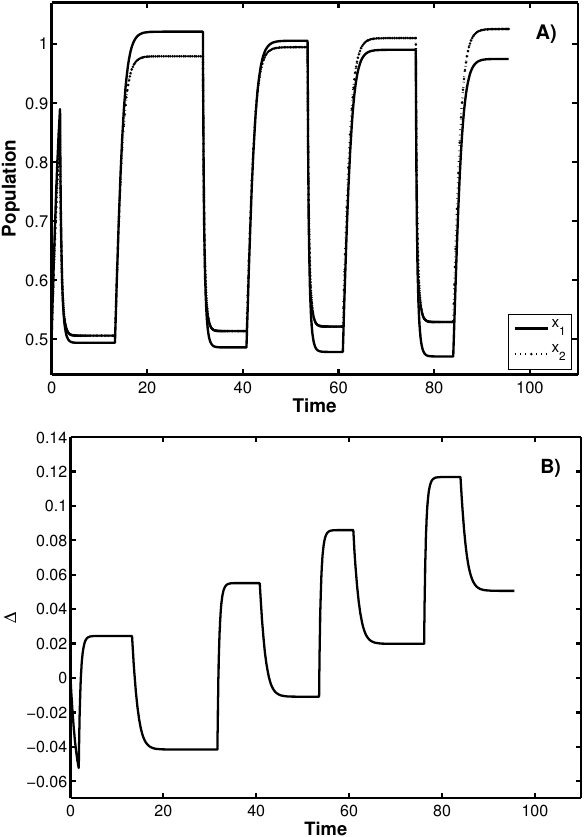}}
\caption{Mutant-ancestor competition for environment fluctuating
  between two states. (A) Time dependence of $x_1$ and $x_2$. (B) Time
  dependence of the logarithm of the population ratio, $\Delta({\cal
    T})$. The two states are characterized by
  $(r_1=2.3, \, r_2=2, \, K_1=K_2=1)$ and $(r_1=1.1, \, r_2=1, \,
  K_1=K_2=2)$. The time spent in each state is uniformly distributed
  between 4 and 10. The equal carrying capacities ensure zero qst
  contribution while the mutant $x_2$ is winning in the long-term due
  to the geometric contribution. The evolutionary pressure is exerted
  only at the very beginning of the residence period in each
  environmental state, and the total population and the mutant
  fraction stay constant for the rest of each phase. Thus the mutant
  ratio drift depends only on the number of switches, but not on the
  duration of the process. \label{fig:fig3} }
\end{figure}

The approach can be extended to a more common model of piecewise
constant environments, see e.g.,~Refs.~\cite{Kussell05,Mustonen08}.
Consider the case of parameters abruptly changing between $m$ sets
indexed by $\mu=1\dots m$,
$\left(r_1^{\mu},r_2^{\mu},K_1^{\mu},K_2^{\mu}\right)$, at possibly
random times $t_a$. The state occupied between $t_{a}$ and $t_{a+1}$
will be denoted by $\mu_a$. We assume that the interval
$(t_{a+1}-t_a)$ is long enough so that the total population $x(t)$
reaches the carrying capacity long before the environment switches
again, that is $1/r^{\mu_i} \ll (t^{a+1}-t^a)$. In this case one can
derive the qst contribution as a sum over all of the environment states
\beq \Delta_{\rm qst}= \sum_a r^{\mu_a} \frac{\delta
  K^{\mu_a}}{K^{\mu_a}} (t^{a+1}-t^a). \eeq At each switch, there is
an extra contribution because $x(t>t_a)$ reaches the value $K^{\mu_a}$
with a delay. That is, from Eq.~\aref{eq:xsol}, we derive: \beq
x(t_a<t<t_{a+1})={K^{\mu_{a}}}\left[{\frac{K^{\mu_{a}}-
      K^{\mu_{a-1}}}{ K^{\mu_{a-1}}}
  \rme^{ r^{\mu_a}(t^a-t) }}+1\right]^{-1}.
\label{eq:solpm}
\eeq Integrating Eq.~\aref{eq:solpm} results in a geometric
contribution after $M$ environment state changes \beq \Delta_{\rm
  geom}({\cal T})=\sum_{a} \left[ \frac{\delta
    r^{\mu_a}(\lambda)}{r^{\mu_a}(\lambda)} -\frac{\delta
    K^{\mu_a}(\lambda)}{K^{\mu_a}(\lambda)}
\right]\log\left[\frac{K^{\mu_a}(\lambda)}{K^{\mu_{a-1}}(\lambda)}
\right].
\label{eq:I2s}
\eeq The fact that Eq.~\aref{eq:I2s} is independent of the actual time
spent in each state and depends only on the sequence of environmental
states is the signature of its geometric nature, illustrated in
Fig.~\ref{fig:fig3}.  Importantly, unlike in the continuos variation
case, Eq.~\aref{eq:I}, $\Delta_{\rm geom}$ in Eq.~\aref{eq:I2s} can
have a finite value even if parameters change only between two states.
Hence it is unclear if the contribution can be interpreted as an
oriented area enclosed by the trajectory in the parameter space.

\section{Continuous stochastic environments}

Often environments change in a continuous but unpredictable way, such
that the typical rate of change is still small. This scenario is
modeled with Gaussian fluctuations of the parameters
\cite{Turelli77,Lande09}. Denoting all parameters with a single symbol
$\gamma_\alpha, \alpha=1,\dots,A$, we generalize our result,
Eq.~\aref{eq:I}, and represent the geometric contribution for randomly
driven Eqs.~(\ref{eq:megm}) as a line integral
\cite{Sinitsyn07a,Sinitsyn07b,Sinitsyn10} 
\beq \Delta_{\rm
  geom}({\cal{T}})=\int_0^{\cal{T}} \!dt\, \sum_{\alpha=1}^A
f_\alpha(\gamma_1(t),\, \dots, \gamma_A(t))\, \dot \gamma_\alpha(t).
\label{eq:gen}
\eeq Here ${\cal{T}}$ is a long time that allows for averaging, and
$f_\alpha$ are some model-dependent functions. Since fluctuations are
small, we expand $f_\alpha$ to the first order in the fluctuations
around the average parameters \beq f_\alpha(\gamma_1(t),\, \dots,
\gamma_A(t))=f_{0\alpha}+ \sum_{\beta=1}^A
\kappa_{\alpha\beta}\gamma_\beta(t).  \eeq Now using suitable
continuity properties of the parameters' trajectory, we transform the
geometric contribution to \beq \Delta_{\rm
  geom}({\cal{T}})=\int_0^{\cal{T}} \! dt \sum_{\alpha=1}^A
\sum_{\beta=\alpha+1}^A
(\kappa_{\beta\alpha}-\kappa_{\alpha\beta})\gamma_\alpha(t) \dot
\gamma_\beta(t) .
\label{eq:con}
\eeq The geometric properties of $\Delta_{\rm geom}$ are clear from
Eq.~\aref{eq:con}: $\Delta_{\rm geom}({\cal{T}})$ depends only on the
length of the parameters' trajectory, is antisymmetric with respect to
time reversals, and is nonzero only if multiple parameter vary
simultaneously and incoherently. Note that Eq.~\aref{eq:con} is valid
only for parameter variations with small (bounded) speeds. Therefore,
if the parameter dynamics, $\gamma_\alpha(t)$, are modeled as
multidimensional Wiener processes, care must be taken to regularize
and properly define the stochastic integrals in
Eqs.~(\ref{eq:gen},~\ref{eq:con}) \cite{Turelli77}.

Equations \aref{eq:gen} and \aref{eq:con} represent a natural
extension of the geometric correction to acyclic trajectories
\cite{Sinitsyn10}. While now the geometric term $\Delta_{\rm
  geom}({\cal{T}})$ is aperiodic, for parameters dynamics with a
stationary distribution of $\gamma_\alpha$ and $\dot\gamma_\alpha$,
$\Delta_{\rm geom}({\cal{T}})$ still has a mean linear dependence on
${\cal{T}}$ for large times: \beq
\lim_{{\cal{T}}\to\infty}\frac{\Delta_{\rm geom}({\cal{T}})}{\cal{T}}=
\sum_{\alpha=1}^A \sum_{\beta=\alpha+1}^A
(\kappa_{\beta\alpha}-\kappa_{\alpha\beta})\left.\frac{d
    C_{\alpha\beta}(t)}{d t} \right|_{t=0},
\label{eq:b}
\eeq where $C_{\alpha\beta}(t) =\avg{\gamma_\alpha(0)
  \gamma_\beta(t)}$ are time dependent correlations of the
environment. Note that the derivatives $\left.\frac{d
    C_{\alpha\beta}(t)}{d t} \right|_{t=0}$ are inversely proportional
to the correlation times of the process. Moreover one can identify the
terms in the rhs of Eq.~\aref{eq:b} as products of the Berry
curvature, $\kappa_{\beta\alpha}-\kappa_{\alpha\beta}$, previously
introduced in the classical and quantum geometric phases literature
\cite{Shapere1988,Sinitsyn07b}, and, for $\alpha\neq\beta$, the rates
of growth of the oriented areas bounded by the process $\left.\frac{d
    C_{\alpha\beta}(t)}{d t} \right|_{t=0}$.

\section{Possible experimental effects}

The existence of geometric corrections to fitness in a time dependent
environment requires that changes in the environment are felt by the
population on multiple time scales. In the model, Eq.~(\ref{eq:mot}),
the immediate change in the growth rates and the delayed effect of the
population reaching the carrying capacity provide these scales, but
other mechanism would work as well. Similar effects will be
encountered in almost any situation when a population responds to
asynchronous changes in multiple external stresses or nutrient
supplies. Therefore, the geometric effects must be considered when
modeling emergence or fixation of new metabolic or stress-resistance
functions in the presence of environmental changes. We suggest that
the {\em relative timing} of fluctuations of extracellular
nutrient/stressor concentrations will affect the relative fitness
advantage of these functions.

\begin{figure}
\centerline{\includegraphics[width=6cm]{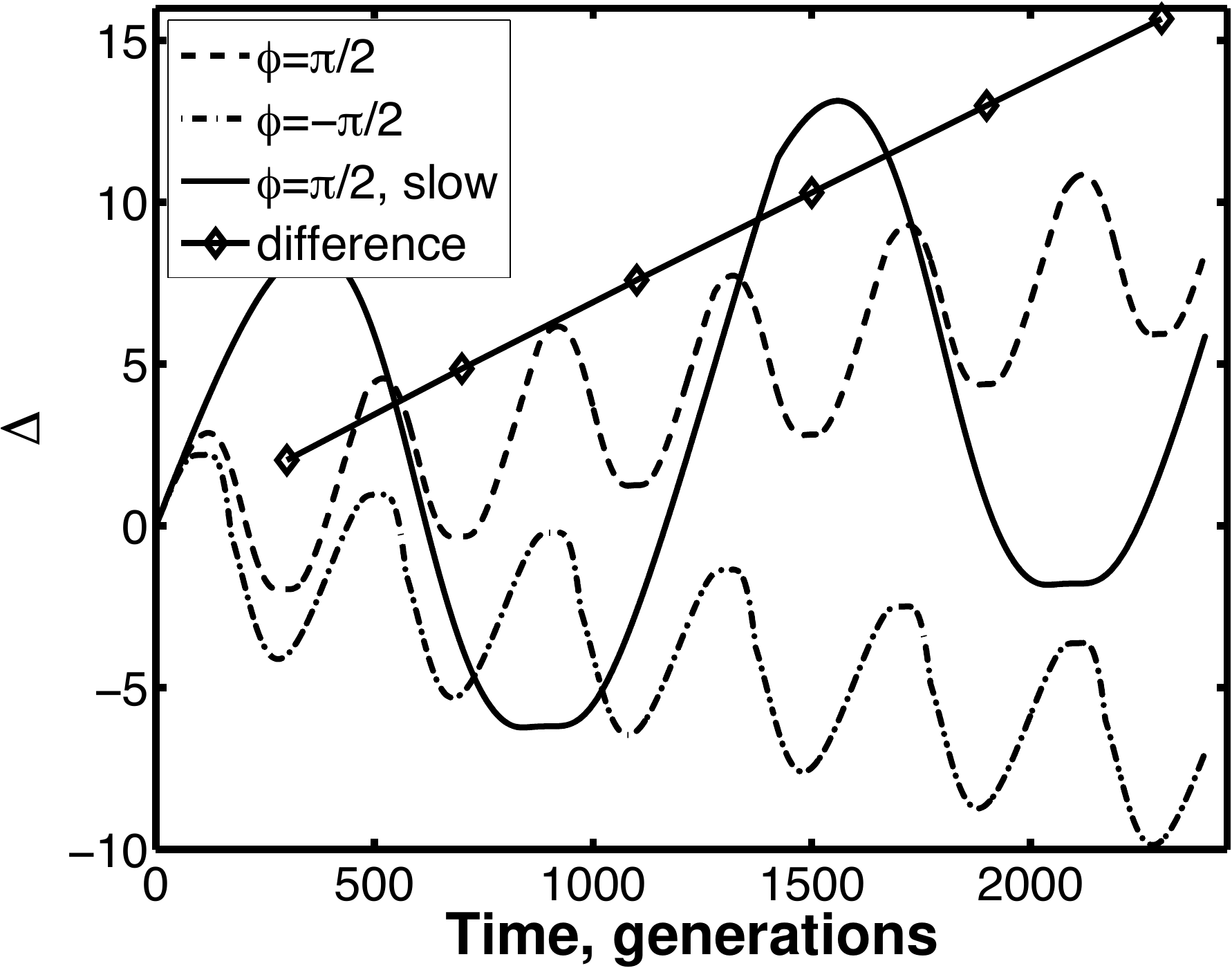}}
\caption{Simulated dynamics of the logarithm of the relative
  population size for two partially antibiotic resistant populations
  competing for the same consumable carbon source in a chemostat. The
  concentration of the antibiotic changes as $A(t)=0.1+0.1\cos(\omega
  t)$ in arbitrary units. The nutrient influx is $1.0+\cos(\omega
  t+\phi$). The nutrient is cleared by the chemostat and consumed by
  both strains in proportion to the population growth, resulting in
  the concentration $\nu(t)$. The growth rate of either population is
  proportional to
  $V\left(1+A/K_A\right)^{-1}\left(1+K_\nu/\nu\right)^{-1}$. $V$ is
  the maximum growth rate, $K_\nu$ is the Monod growth constant, and
  $K_A$ is related to the minimal inhibitory concentration for the
  antibiotic. $K_A$ for the more resistant strain is 14\% higher than
  for the less resistant one, but its $V$ is 5\% smaller to account
  for the cost of resistance \cite{Andersson2010}. The numbers are
  chosen such that the average growth for very slow environmental
  changes (solid line) is almost the same for both strains. Depending
  on the phase $\phi$ (dashed and dash-dotted lines), either the
  resistant or the non-resistant strain has the higher growth rate and
  will eventually take over the population. The ``difference'' line
  shows the nearly linear difference between the strain fractions for
  the two opposite environmental trajectories.\label{fig:anti} }
\end{figure}

Of a particular interest is emergence of antibiotic resistance in
bacteria. Mutations conferring antibiotic resistance often decrease
ability of cells to grow in the absence of antibiotics, but provide a
growth advantage in their presence \cite{Andersson2010}. At the same
time, delivery of antibiotics is hardly ever uniform, and nutrient
supplies also fluctuate. Focusing for simplicity on periodic nutrient
and antibiotics concentration changes, we see that the {\em time
  delay}, or the {\em phase
  lag}, between the changing concentrations will join their amplitudes
and the period in selecting whether a resistant strain will fix or
not. We illustrate this in Fig.~\ref{fig:anti}: depending on the phase
difference between the nutrient and the antibiotic influx, either the
resistant or the faster growing bacterium will be selected for. A
robust prediction of our theory is that the difference in the
logarithmic fractional population changes between an environmental
trajectory and its time reversed counterpart will grow almost linearly
in time with the number of periods. We emphasize that the effect is
different from episodic selection \cite{Johnsen2009}, where only
frequencies and magnitudes of antibiotic selection episodes determine
fixation of the resistant strain.

Another experimental system where our predictions can be important is
evolution of a metabolic pathway corresponding to a new metabolite,
when both the old and the new metabolite concentrations change in
time. In such a case, one would need to take into account possible
effects of catabolite repression and di-auxic growth in addition to
instantaneous effects of metabolite concentrations on the birth/death
rates. Nevertheless we expect that careful modeling of these effects
will also uncover the fitness sensitivity to the timing of pathway
activation. 

An important characteristic of the geometric effect is that it is much
harder to be observed in typical serial dilution experiments,
especially when the environment changes are only imposed at the
dilution points. Such experimental protocols will miss important
effects that may be relevant for wild-type conditions.

\section{Discussions}

Fixation dynamics of mutants in a large class of mathematical models
is governed by a single effective parameter, the selection rate,
obtained as a time average of the instantaneous growth rate difference
between the mutant and the ancestral population. In population
dynamics with symmetric competition, and in the limit of small
differences between the mutant and the ancestor, the total population
size is decoupled from changes in the population composition. Instead
the total population enters the fixation dynamics only as a time
dependent parameter. Then the population growth rates and the
selection coefficient depend on the interplay between the time scales
of the population dynamics and the environmental fluctuations. For
infinite separation between the time scales, the selection depends
only on values of environmental parameters. More specifically, here
the fitness difference can be expressed as a function of growth rates
and carrying capacities averaged over all of the environmental states
and independent of the period of the fluctuations. Nonetheless, due to
the non-linear dependence of the growth rates on the environmental
parameters, the average fitness difference is not necessarily the same
as the fitness difference for the average environment.

This quasi-steady state approximation breaks down for faster
environmental changes. The mutant fraction dynamics is now dependent
not only of the period of environmental changes but also on the
sequence of successive environmental states. In particular the first
non-adiabatic correction is always anti-symmetric with respect to time
reversals, and it is geometric in nature. As long as the fluctuations
in the parameters are large, this non-adiabatic correction can be of
the same order of magnitude in the birth and death rates variation as
the qst contribution to the fitness difference. The geometric nature
of this term constrains the effect that environment fluctuations can
have on fitness differences. Indeed, as other geometric contributions
\cite{Shapere1988,Sinitsyn09}, this effect is independent of the
instantaneous speed of variation of parameters. In ecological terms,
this implies that the geometric contribution to the mutant ratio drift
does not depend on how fast the environment changes, but only on the
sequence of environmental states. We illustrate this in
Figs.~\ref{fig1},~\ref{fig:fig3}. Further, we note that the mutant
fraction drift, $\Delta$, can be seen as a line integral in the
parameter space, cf.~Eq.~ \aref{eq:I}. This implies that only
multidimensional and off-phase parameter variations can give nonzero
long-term contributions to the population dynamics. 

For the results derived in this work, the assumption of an oscillatory
environment is not essential. Our conclusions, and the concept of
geometric phase in general, are valid for non-cyclic environment
dynamics \cite{Sinitsyn10}. Typically such dynamics is represented
with a Gaussian and, in general, uncorrelated noise
\cite{Heckel80,Gillespie94,Lande09}. While a detailed extension of the
present results to random trajectories is beyond the scope of this
paper, we have shown here that the geometric contribution to the
selection coefficient is present generically if and only if the
population dynamics contains multiple correlated parameters driven by
a colored noise, cf.~Eq.~(\ref{eq:b}).

In this article we have focused on deterministic population dynamics
with small parameter differences among the competing species, which is
equivalent to frequency independent selection. We expect that a
similar geometric phase contribution to the fixation dynamics is
present in stochastic Fisher-Wright type models, as well as models
that exhibit various frequency and density dependent selection
effects.

The results in Eq.~(\ref{eq:I}) allow us to make a conclusion that is
independent of the exact variation of the parameters and the exact
details of the model. Namely, for clones with the same mean fitness,
{\em the clone that has a higher growth rate when the environment is
  abundant (increasing carrying capacity) will have a selective advantage}
over the clone that performs well when the carrying capacity
decreases. This is important during acquisition of new metabolic or
stress-response functions, as discussed above. Further, in the case of
the long-term {\em E.\ coli} evolution experiment \cite{Lenski11}, we
point out that unless mutations manifest themselves in a positive way
during the exponential growth phase following a serial dilution, daily
variability of the environment would make it harder for mutations to
fixate even without stochastic effects associated with the dilution
bottlenecks.

We conclude with an observation that species with the fitness
advantage in the average environment, with the average fitness
advantage over all environments, and with the average fitness
advantage for a particular time course of the environment are not
necessarily the same species. In particular, a naively deleterious
mutation can fixate in a population due to these temporal effects. We
believe this to hold true independently of many of the simplifying
assumptions of our toy model.

\begin{acknowledgments}
  We thank B Levin, J Otwinowski, M Tchernookov, and N Sinitsyn for
  important discussions that have shaped this work. We are
  particularly grateful to B Levin for his insightful critique of the
  manuscript and the approach.
\end{acknowledgments}

\end{document}